\documentclass[10pt]{article}

\usepackage[superscript,sort]{cite}

\usepackage{ifthen,ifpdf}
\ifpdf
	\usepackage[pdftex]{hyperref}	\hypersetup{colorlinks=true,linkcolor=black,citecolor=black,urlcolor=blue,backref=page,bookmarks=true,breaklinks=true,plainpages=false }
	\usepackage[pdftex]{graphicx}
	\usepackage[usenames,pdftex]{color}
\else
	\usepackage[colorlinks=true,breaklinks=true,plainpages=false]{hyperref}
	\usepackage{graphicx}
	\usepackage[usenames]{color}
\fi

\usepackage{amsthm,amscd,amsxtra,amsfonts,amsmath,amssymb,multirow}
\usepackage{wrapfig}
\usepackage[footnotesize]{caption}
\usepackage[tiny,compact]{titlesec}
\usepackage[textwidth=0.8in,textsize=footnotesize]{todonotes}
\usepackage{algorithm,algorithmic,extarrows}

\begin{document}

\title{Multiscale persistent functions for biomolecular structure characterization }

\author{
Kelin Xia$^1$ \footnote{ Address correspondences  to Kelin Xia. E-mail:xiakelin@ntu.edu.sg}
and
Zhiming Li$^2$ and
Lin Mu$^3$  \\
$^1$Division of Mathematical Sciences, School of Physical and Mathematical Sciences, \\
Nanyang Technological University, Singapore 637371\\
$^2$Key Laboratory of Quark and Lepton Physics (MOE) and Institute of Particle Physics\\
Central China Normal University, Wuhan 430079, China \\
$^3$ Computer Science and Mathematics Division\\
Oak Ridge National Laboratory, Oak Ridge, TN, 37831, USA
}

\date{\today}
\maketitle

\begin{abstract}
In this paper, we introduce multiscale persistent functions for biomolecular structure characterization. The essential idea is to combine our multiscale rigidity functions with persistent homology analysis, so as to construct a series of multiscale persistent functions, particularly multiscale persistent entropies, for structure characterization. To clarify the fundamental idea of our method, the multiscale persistent entropy model is discussed in great detail. Mathematically, unlike the previous persistent entropy or topological entropy\cite{Chintakunta:2015entropy,Merelli:2015topological,Rucco:2016}, a special resolution parameter is incorporated into our model. Various scales can be achieved by tuning its value. Physically, our multiscale persistent entropy can be used in conformation entropy evaluation. More specifically, it is found that our method incorporates in it a natural classification scheme. This is achieved through a density filtration of a multiscale rigidity function built from bond and/or dihedral angle distributions. To further validate our model, a systematical comparison with the traditional entropy evaluation model is done. It is found that our model is able to preserve the intrinsic topological features of biomolecular data much better than traditional approaches, particularly for resolutions in the mediate range. Moreover, our method can be successfully used in protein classification. For a test database with around nine hundred proteins, a clear separation between all-alpha and all-beta proteins can be achieved, using only the dihedral and pseudo-bond angle information. The persistent entropy values of mixed-alpha-and-beta proteins are also found to be in the middle region with just a few cases overlapped with the other two categories. Finally, a special protein structure index (PSI) is proposed, for the first time, to describe the ``regularity" of protein structures. Basically, a protein structure is deemed as regular if it has a consistent, uniformed, and orderly configuration. Our PSI model is tested on a database of 110 proteins, we find that structures with larger portions of loops and intrinsically disorder regions are always associated with larger PSI, meaning an irregular configuration. While proteins with larger portions of secondary structures, i.e., alpha-helix or beta-sheet, have smaller PSI. Essentially, PSI can be used to describe the ``regularity" information in any systems.
\end{abstract}

\newpage



\section{Introduction}

A fundamental ingredient of almost every biological process is molecular recognition, which is widely observed in interaction systems like protein-protein, receptor-ligand, antigen-antibody, DNA-protein, RNA-ribosome, etc\cite{Gellman:1997,Brooijmans:2003,Janin:2013}. Understanding the basis for molecular recognition requires the full characterization of binding process which involves noncovalent bonding effect and biomolecular thermodynamics. Among all the factors, biomolecular conformation energy is of great importance\cite{Frederick:2007,Marlow:2010role}. It is found that changes in protein conformational entropy can contribute significantly to the free energy of protein-ligand association\cite{Frederick:2007}. Internal dynamics of the protein calmodulin varies significantly on binding a variety of target domains. The apparent change in the corresponding conformational entropy is linearly related to the change in the overall binding entropy. Also,
conformational entropy of protein side-chain is a major effect in the energetics of folding. The entropy of heterogeneous random coil or denatured proteins is significantly higher than that of the folded native state tertiary structure. In particular, the conformational entropy of the amino acid side chains in a protein is thought to be a major contributor to the energetic stabilization of the denatured state and thus a barrier to protein folding. The reduction in the number of accessible main chain and side-chain conformation when a protein folds into a compact globule yields an unfavourable entropic effect. This reduction in conformational entropy counters the hydrophobic effect favoring the folded state and in part explains the marginal stability of most globular proteins.

Even thought biomolecular conformation energy is a key property to understand a wide variety of physical, chemical, and biochemical phenomena, its evaluation or calculation is very challenging both experimentally and theoretically. Only recently, NMR relaxation methods for characterizing thermal motions on the picosecond-nanosecond (ps-ns) timescale is developed and the resulting order parameters is used as a proxy for conformational entropy evaluation\cite{Frederick:2007,Trbovic:2008,Sapienza:2010}. Atomic force microscopy (AFM) for unfolding has great potential in measuring the backbone conformational entropy for protein folding\cite{Thompson:2002}. Neutron spectroscopy is also used to elucidate the role of conformational entropy upon thermal unfolding by observing the picosecond motions, which are dominated by side-chain reorientation and segmental movements of flexible polypeptide backbone regions\cite{Fitter:2003}.

The computational estimation of conformation energy is a long-standing problem and one of challenges in computational chemistry\cite{Baron:2009}. Generally, the calculation of conformational entropy requires a fully understanding of biomolecular configuration spaces. Various simulation techniques, including harmonic analysis, molecular dynamics (MD), Monte Carlo simulation, normal mode analysis and so on, are all employed to explore the configuration spaces\cite{Hagler:1979,Karplus:1981}. Traditionally, the simulation data is processed by the quasiharmonic analyses\cite{Baron:2009}. More specifically, a special variance matrix for the biomolecular conformations can be defined and analyzed with principal component analysis (PCA). Conformation entropies are then evaluated from the eigenvalue and eigenvector information from PCA. 
However, for macromolecules, quasiharmonic analyse can be computationally demanding. And the employment of the Cartesian coordinates representation makes it very inefficient for bond rotation characterization\cite{Das:2013}. Moreover, it is found that, for biomolecular systems with multiple occupied energy wells, quasiharmonic approximation tends to overestimate their configurational entropy, and with Cartesian coordinates, the errors tend to be magnified\cite{Chang:2005}.

Characterization of biomolecular conformations by some structural parameters has been proposed to estimate the conformational entropy\cite{Trbovic:2008}. Particularly, backbone dihedral angles and side chain rotamers are widely used structural measurements for protein conformational entropy evaluation. The entropy value depends on the probability of the occupancy of the particular structure. By the assumption that amino acids distributions in the native state of proteins are comparable to that found in the denatured state, Stites and Pranata\cite{Stites:1995} proposed a way to evaluate the relative conformation entropies for different amino acids. They analyze the preferred distribution of amino acid residues by systematically studying about twelve thousand residues from sixty-one nonhomologous and high-resolution protein crystal structures.  A Ramachandran plot of $(\phi, \psi)$ angles for various amino acid residues are obtained and the conformation entropies are evaluated through a discretization of angle distribution by an uniformed grid size.  The dihedral angle and sid-chain rotamer based parameterization has been widely used in biomolecular conformation entropy estimation\cite{Doig:1995side,Brady:1997entropy,Zhong:2006dihedral,Zhang:2008RNA,Baruah:2015}.

The above-mentioned dihedral-angle based entropy evaluation method usually involves a discretization of angle distributions and it is found that the calculated entropy values are sensitive to the grid size\cite{Stites:1995,Trbovic:2008}. It is simply true that when the grid is very fine or dense, angles with similar values can be classified into different categories. In contrast, when the grid is very coarse, even angles with large variation can still be classified into a same category. Therefore, dramatically different entropy values can be obtained from the same data under different discretization. An example can be found in Figure \ref{fig:entropy_problem}. However, it has also been pointed out that correlation coefficients between entropies computed from different meshes are very high, suggesting that mesh-related bias does not systematically alter relative entropy values\cite{Stites:1995,Trbovic:2008}. But this consistence is highly related to the studied systems. Simply speaking, for each Ramachandran plot, only the same type of amino acide is considered thus the angle is highly concentrated in a particulary area instead of scattering over a wide range. Researchers have realized that a lack of a robust classification poses challenge to a rigorous estimation of the entropies. Recently, a k-mean clustering method is proposed to deliver an optimized discrete k-state classification model\cite{Zhang:2008RNA}. In this method, the distribution of torsional angles are naturally classified in to $k$ clusters with irregular boundaries. In this way, it achieves an optimized classification and a high Silhouette value, indicating very good quality of clustering, is obtained. In this paper, we propose a multiscale persistent function for biomolecular structure characterization and conformation entropy evaluation. Our multiscale persistent function is based on two major methods, i.e., persistent homology and flexibility rigidity index (FRI).

Persistent homology is deeply rooted in algebraic topology but finds great potential in the simplification of complex data \cite{Frosini:1999,Robins:1999, Edelsbrunner:2002,Zomorodian:2005}. Unlike the traditional topological method, persistent homology provides a multiscale topological representation. It is able to embed a geometric measurement into topological invariants and provides a bridge between geometry and topology. The essence of persistent homology is filtration. By the consistent variation of a certain filtration parameter, a series of topological spaces from various scales are generated. During the filtration, topological invariants, i.e., Betti numbers, will be generated and further continue or persist for some time. This life length or persistent time gives a relative geometric measurement of the associated topological properties. Persistent homology finds great success in topological characterization identification and analysis (CIA) and has been used in a variety of fields, including shape recognition \cite{DiFabio:2011},network structure \cite{Silva:2005,LeeH:2012,Horak:2009},image analysis \cite{Carlsson:2008,Pachauri:2011,Singh:2008,Bendich:2010,Frosini:2013}, data analysis \cite{Carlsson:2009,Niyogi:2011,BeiWang:2011,Rieck:2012,XuLiu:2012},chaotic dynamics verification \cite{Mischaikow:1999,Kaczynski:2004}, computer vision \cite{Singh:2008} and computational biology \cite{Kasson:2007,YaoY:2009, Gameiro:2013}. Recently, we have introduced persistent homology for structure characterization and mathematical modeling of fullerene molecules, proteins and other biomolecules \cite{KLXia:2014c, KLXia:2015a,BaoWang:2016a}. Consistent barcode patterns are extracted and defined as molecular topological fingerprint (MTF), which is used in the analysis of protein flexibility and protein folding \cite{KLXia:2014c}. We have also developed multiresolution and multidimensional persistence \cite{KLXia:2015c,KLXia:2015b}.

Flexibility rigidity index is originally proposed to study the flexibility and dynamic modes of biomolecules, particularly for the prediction of Debye-Walter factors or B-factors\cite{KLXia:2013d,Opron:2014,Opron:2015communication,Opron:2014b, Xia:2015multiscale,Nguyen:2016generalized}. The basic assumptions of the FRI method is that biomolecular structure is the equilibrium state determined by interactions within the structure. Simply speaking, if two atoms or residues are close to each other,``general interaction" between them will be relatively strong. If they are far from each other, ``general interaction" will be relatively week. In this way, a distance-based monotonically-decreasing function is chosen to describe the particle-particle interactions\cite{Xia:2014molecular}. Here a particle can be an atom, a residue or a certain domain. A rigidity index is defined on each particle as the summation all of interactions between this particular particle and all the others. And flexibility index defined as the reciprocal of rigidity index is found to be linearly proportional to the experimental B-factors. More interestingly, our flexibility index can also be viewed as the closeness centrality in graph theory. The physical insight is that atoms located in the center will tend to have more neighbors thus more rigid, while atoms located far from the center will have less companions thus less rigid and more flexible. A rigidity function can be constructed to simulate the biomolecular density \cite{Nguyen:2016generalized}.

The essential idea of our multiscale persistent functions is the combination of persistent homology analysis with our multiscale rigidity function, to construct a series of multiscale persistent functions, particularly multiscale persistent entropies, for structure characterization. Unlike the previous persistent entropy or topological entropy\cite{Chintakunta:2015entropy,Merelli:2015topological,Rucco:2016}, we incorporate a resolution parameter into our rigidity function and delivers a topology based multiscale entropy model. Based on the density filtration, our method incorporates in it a classification scheme. The classification results are reflected in $\beta_0$ barcodes, and are further used in persistent entropy\cite{Chintakunta:2015entropy,Merelli:2015topological,Rucco:2016} calculation. We have systematically compared our entropy model with the traditional model with fixed grid sizes and found that similar results can only be obtained in two extreme situations, i.e., the grid size and resolution parameter value are very small or very large. In the middle range, different results are obtained as our classification is different. Our method is tested for protein structure classification. We found that it is able to deliver a very nice classification of all-alpha, all-beta and mixed alpha and beta proteins. The all-alpha and all-beta proteins are dramatically different and can be clearly separated by their conformation entropies. The mixed alpha and beta proteins have both alpha-helix and beta-sheet secondary structures. Their entropies are largely distributed between the border of all-alpha and all-beta entropies. The results are comparable with our previous ones from machine learning method \cite{ZXCang:2015}. More interestingly, based on our multiscale persistent entropy, a protein structure index (PSI) is proposed to describe the ``regularity" of protein structures. The essential idea of PSI is to evaluate disorderliness in the angle distributions. Simply speaking, for a highly ``regular" structure element like alpha-helix, its dihedral and bond angles are very consistent thus contribute very little to the total entropy. Loops and intrinsically disorder regions are extremely ``irregular" in terms of dihedral and bond angles and tend to contribute a large weight in the total entropy. In this way, through the entropy evaluation, PSI is able to provide a new way of structure regularity characterization. Essentially, PSI can be used to describe the regularity information in any systems.

The paper is organized as following, Section \ref{sec:theory} is devoted to method introduction. We give a brief introduction of conformation entropy in Section \ref{sec:conformation_entropy}. In this section, the discrete representation of protein structure and angle-based conformation entropy evaluation are discussed. Multiscale rigidity function is introduced in Section \ref{sec:Multiscale_rigidity_function}. Rigidity function is essentially a continuous version of rigidity index, and our multiscale representation is achieved through tuning a resolution parameter. The persistent homology analysis, is reviewed in Section \ref{sec:PHA}, which includes the basic theory of simplicial complex, filtration, complex construction and persistence. Section \ref{sec:MPH} is devoted for the multiscale persistent homology and multiscale persistent functions. We propose a density based filtration over the multiscale rigidity function and define various functions on the persistent barcodes. The multiscale persistent entropy is discussed with great details. Finally, the application of our model can be found in Section \ref{sec:application}. A classification of various protein structures including all-alpha, all-beta, mixed alpha and beta,  are studied in Section \ref{sec:classification}. Section \ref{sec:protein_index} is devoted for the description of protein structure index (PSI). The paper ends with a conclusion.

\section{Method and algorithm}\label{sec:theory}
Entropy is proposed for the characterization of disorderlyness of a system\cite{Karplus:1981,Brady:1997entropy}. It measures the freedom of a system to evolve into various potential configurations. Entropy is the key property to understand a wide variety of physical, chemical, and biochemical phenomena. It plays very important roles in various biomolecular structures and interactions including protein folding, protein-protein interaction, protein-ligand binding, chromosome configuration, DNA translation and transcription, etc. For instance, the folding of a single peptide chain into a well-defined native structure is greatly facilitated by the reduction of its conformation entropy. Therefore, conformation entropy calculation is of essential importance in computational chemistry, biophysics and biochemistry.

The evaluation of the entropy necessitates a characterization of biomolecular configuration spaces. Due to the limitation of Cartesian grid based representation, structural parameters, particulary dihedral angles and bond anlges, are usually employed for structure description. Their distributions can be derived from various computational methods, including molecular dynamics, Monte Carlo simulation, normal mode analysis, etc. Various microstates can be obtained by the discretization of the angle distribution, usually through a equal-spacing grid. And conformation entropy can be evaluated by using the classic  Shannon entropy form. Even though it is suggested that the relative entropy is consistent in this procedural, a lack of a robust classification still poses challenge to a rigorous estimation of the entropie. In this section, multiscale persistent analysis and multiscal persistent functions for biomolecular structure characterization and conformation entropy evaluation are introduced. This methodology is founded on two major methods, i.e., persistent homology and flexibility rigidity index. The great details will be given in the following sections. But before that, a brief review of the conformation entropy is given firstly.

\subsection{Conformational entropy}\label{sec:conformation_entropy}
In physics, the second law of thermodynamics states that the total entropy of an isolated system always increases. This law also leads to the Clausius thermodynamic definition of entropy as following:
\begin{eqnarray}\label{eq:entropy_second}
 \Delta S = \int \frac{\delta Q}{T},
\end{eqnarray}
here $\delta Q$ is the amount of heat absorbed by the system.
For a real biomolecular system, this definition of entropy is too abstract and hard to implement. Normally, a direct calculation of entropy is based on the Boltzmann-Gibbs-Planck (BGP) expression
\begin{eqnarray}\label{eq:entropy_integration}
S=-K_B \int p({\bf r})ln( p({\bf r}) ) d {\bf q},
\end{eqnarray}
where $K_B$ is the Boltzmann constant and $p({\bf r})$ is the probability distribution function. If the potential function of a biomolecule can be modeled as $U(\bf r)$, probability function will be linearly proportional to $e^{-\frac{U(\bf r)}{K_BT}}$, i.e, $p({\bf r}) \propto  e^{-\frac{U(\bf r)}{K_BT}}$. However, Eq.\ref{eq:entropy_integration} brings challenges in practical biomolecular simulation. This is largely due to the enormous complexity of conformation space and potential function for biomolecules.

To be able to evaluate the entropy, researchers adopt a fundamental assumption that entropy can be divided into several components\cite{Brady:1997entropy}.
In this way, protein folding and binding entropy can be divided into protein contribution $\Delta S_p$ and solvent contribution $\Delta S_s$. Further, protein contribution can be subdivided into conformation entropy $\Delta S_c$, rotational entropy $\Delta S_r$ and translation entropy $\Delta S_t$.

\begin{figure}
\begin{center}
\begin{tabular}{c}
\includegraphics[width=0.8\textwidth]{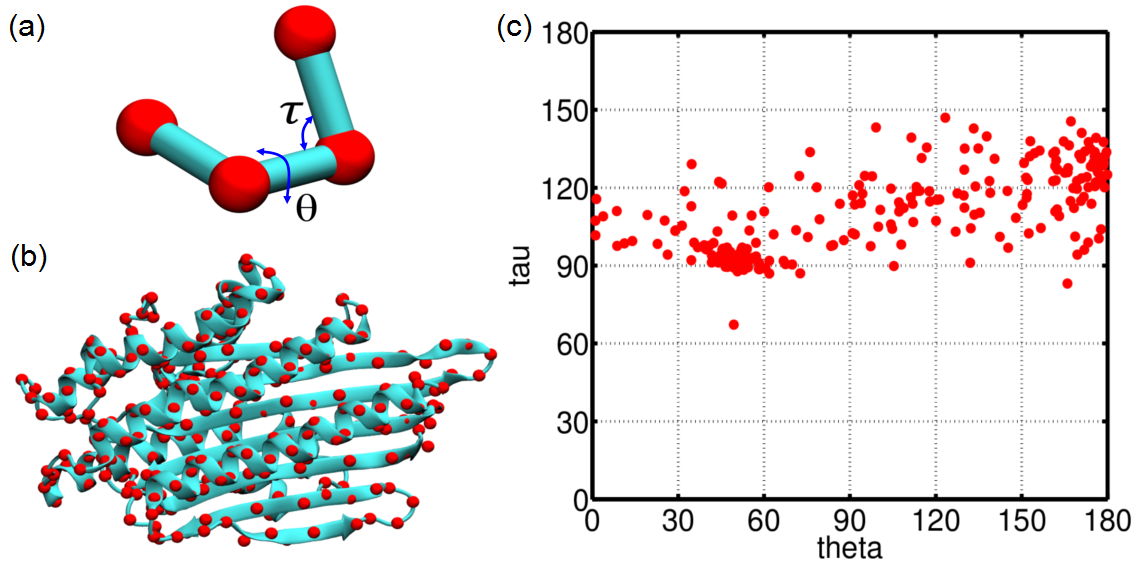}
\end{tabular}
\end{center}
\caption{The illustration of $\theta - \tau$ angle representation of the protein structure. The coarse-grained model with each residue represented by its $C_{\alpha}$ atom (red ball) is used. The dihedral angle is rescale to [0, 180] as the parallel and antiparallel regarded as the same in our protein structure index definition. (a) The angle $\theta$ is the psudo-bond angle between adjacent three $C_{\alpha}$ atoms and angle $\tau$ is the dihedral angle formed by four $C_{\alpha}$ atoms. (b) The coarse-grained model of protein 1VJU (chain A). (c) The $\theta - \tau$ angle distribution of protein 1VJU (chain A).
}
\label{fig:1vju_psi_phi}
\end{figure}

\paragraph{Conformation representation}
In order to quantitatively evaluating the biomolecular conformation entropy, we need to build up models to characterize biomolecular conformation spaces. Generally, dihedral angle models are employed. In these models, biomolecular structures are parameterized by their backbone dihedral angle $\phi$ and $\psi$, and side chain rotametric angles $\chi$. And the probability distribution function can be expressed as $P(\phi,\psi,\chi)$  in this representation. More specifically, three unique dihedral angles can be found on the protein backbone (or protein peptide chain) and all of them are formed between the adjacent four backbone atoms. Dihedral angle $\phi$ is formed between atoms $\{C, N, C_{\alpha}, C\}$, angle $\psi$ is between atoms $\{ N, C_{\alpha}, C, N \}$ and angle $\omega$ is from atoms $\{C_{\alpha}, C, N, C_{\alpha}\}$. Due to the partial-double-bond character, dihedral $\omega$ is within a peptide plannar and its value is normally $180^o$. In this way, each residue can be associated with a pair of  $\phi$ and $\psi$ angles, and protein backbone configuration can be characterized by a two dimensional vector composed of ($\phi$, $\psi$) angles. And Ramachandran plot, a ($\phi$, $\psi$) angle distribution graph, is commonly used to visualized energetically favorable regions.


However, for macromolecules with large number of amino acids, the ($\phi$, $\psi$) representation can be computationally inefficient. To reduce the complexity, many coarse-grained models are used. The most common one is the $C_{\alpha}$ model, in which a whole amino acid residue is represented by its $C_{\alpha}$ atoms. Correspondingly, the backbone configuration can be characterized by virtual dihedral angle $\theta$ and virtual bond angle $\tau$\cite{Levitt:1975computer, Korkut:2013}. To be more specific, virtual dihedral $\theta$ is evaluated from  four consecutive $C_{\alpha}$ atoms and virtual bond angle $\tau$ is evaluated from three consecutive $C_{\alpha}$ atoms. Figure \ref{fig:1vju_psi_phi}(a) illustrates the geometric meaning of angle $\theta$ and $\tau$. In this representation, the $C_{\alpha}$ backbones of proteins are specified with ($\theta$, $\tau$) virtual angles. In analogy to the Ramachandran plot, we can use the distribution of ($\theta$, $\tau$) angles to explore structure properties. Figure \ref{fig:1vju_psi_phi}(b) and (c) illustrates a ($\theta$, $\tau$) angle distributions for protein 1VJU.
With the angle distribution representation, biomolecular structures including protein, DNA, RNA, etc, can be quantitatively characterized. This is essential to the conformation entropy evaluation.

\paragraph{Conformation entropy evaluation}\label{sec:density}
\begin{figure}
\begin{center}
\begin{tabular}{c}
\includegraphics[width=0.9\textwidth]{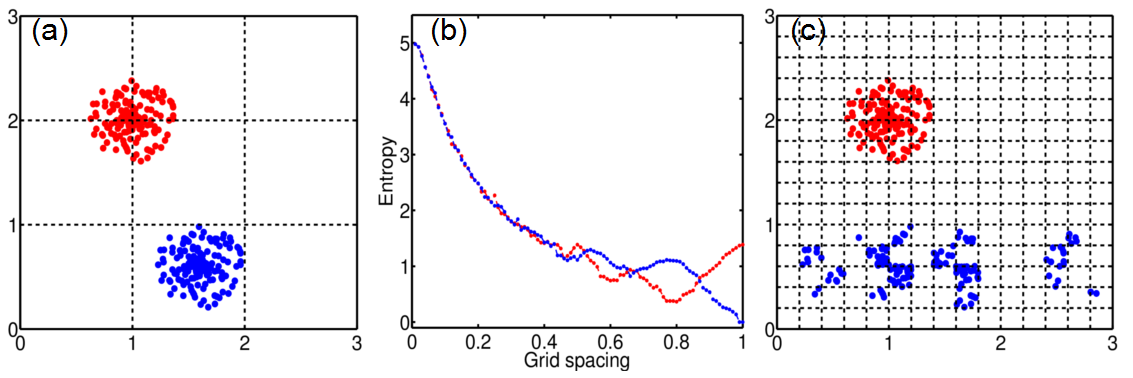}
\end{tabular}
\end{center}
\caption{Illustration of limitation of traditional entropy calculation. (a)Two identical data colored by red and blue has dramatically different entropy values in coarse grid.
(b)The calculated entropy values for two identical data in various mesh spacing. The entropy curves are colored in the same way as the corresponding data sets.(c) The same entropy may represent dramatically different data distributions in a refined mesh.}
\label{fig:entropy_problem}
\end{figure}

As stated above, the configuration spaces and probability distribution function can be parameterized by ($\phi$,$\psi$) or ($\theta$, $\tau$) angles. In conformation entropy evaluation, a discretization of the distribution function is usually done by the subdivision angle distribution into various regions. And a discrete conformation entropy formula can be employed, which is the widely-used Shannon entropy:
\begin{eqnarray}\label{eq:entropy_sum}
S=-\sum  p_i log( p_i)
\end{eqnarray}
here $p_i$ is the probability of the system being in state $i$, and the sum is taken over all possible states of the system. It should be noticed that the Boltzmann constant is no longer exist in this definition, as Shannon entropy is entropy of information.

However, the partition of the conformations into different states is not rigorous and highly depends on the size of the states. Generally, in order to quantitatively assessing the complexity of the models, the angle distributions are usually divided into equally spaced bins and the probability of the state is calculated in each bin. In this way, the size of bin matters a lot in the final value of the conformation entropy. Figure \ref{fig:entropy_problem} demonstrates the great importance of bin size. To evaluate the entropy, the whole region is divided into equally spaced grids. The probability $p_i$ can be evaluated by counting the number of ($\theta$, $\tau$) points in $i$-th grid and dividing it by the total number of points. The domain subdivision procedure has no common standard and is an very subtle issue. For instance, we have two identical data
colored by red and blue and located in different regions of the angle graph. First, a coarse regular 3*3 mesh is used in Figure \ref{fig:entropy_problem}(a). With this discretization, two data sets have dramatically different entropy results, i.e., one is 0.0 and the other is about 1.39. Then, we employ a mesh refinement and subdivide each grid equally into two grids. This time the entropy values are very close, i.e., one is 1.11 and the other is 1.39. To get a whole picture of the relationship between entropy and mesh size, various grad spacings are used and Figure \ref{fig:entropy_problem}(b) shows the corresponding entropy values. To avoid confusion, the entropy points are colored in the same way as the corresponding data. It can be seen that as the mesh grid decreases, two data sets begin to adopt the similar entropy value, which however is not a constant but keeps increasing.  On the other hand, even under a refined mesh, the same entropy may represent dramatically different distributions. As illustrated in Figure \ref{fig:entropy_problem}(c), the grid spacing is 0.2 and the data in blue color is generated by redistributing the grid boxes with red points. As the distribution of data points in grids is not changed, the total conformational entropy is the same.

\begin{figure}
\begin{center}
\begin{tabular}{c}
\includegraphics[width=0.95\textwidth]{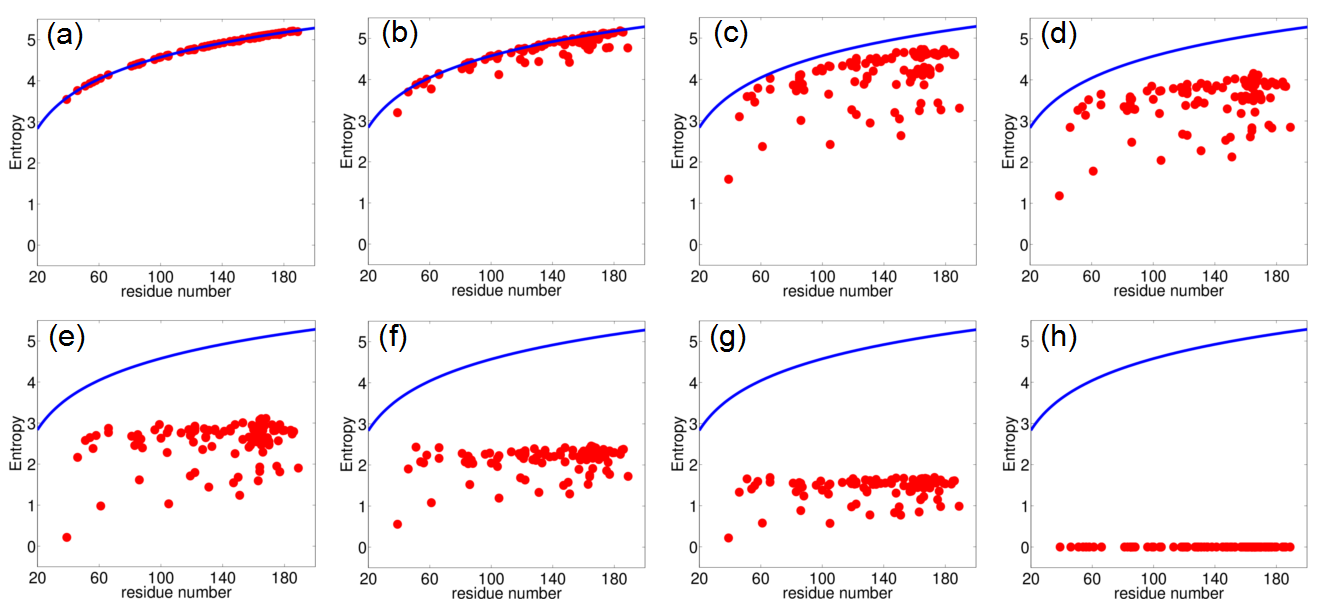}
\end{tabular}
\end{center}
\caption{The illustration of entropy values for the protein set in different mesh sizes. The grid sizes from (a) to (h) are 
900*900, 180*180, 33*33, 18*18, 9*9, 6*6, 3*3 and 1*1, respectively. The blue line represents the function $log(N)$ with $N$ the residue number.
}
\label{fig:entropy_grid}
\end{figure}

To further illustrate the great importance of grid spacing in biomolecular conformational entropy evaluation. We carefully choose a data set with 110 proteins. The protein IDs are listed in Table \ref{tab:protein_index}. All chosen proteins have resolutions smaller than 1.5 \AA~. We process these structures by removing the extra chains in them if any, so that all used data has only one chain. Figure \ref{fig:entropy_grid} demonstrates the relation between the calculated conformational entropy and grid spacing. It can be seen that when the grid spacing is small enough, the conformation entropy converges to $ln(N_{res}-3)$ with $N_{res}$ the number of residues. We need to minus three in the entropy results as we have only $N_{res}-3)$ dihedral angles on the protein backbone. When grid spacing is large enough to incorporate all the points in a single grid, the entropy goes to zero. The blue curve in the each subfigure of Figure \ref{fig:entropy_grid} is for function $ln(x)$.

Essentially, the discretization of the configuration spaces is a classification problem. However, traditional discretization method with a regular mesh is too coarse to capture the inner structures, patterns and features of the configuration spaces. To incorporate more data distribution information into discretization,  the K-mean clustering algorithm is employ\cite{Zhang:2008RNA}. In this method, all angle points are classified into several clusters or configuration states in a way that the summation of a distance function between each point to its cluster center is minimized. State differently, it is clustered based on the relation of each point to its neighbors. Inspired by it, we proposed a multiscale persistent entropy algorithm. Simply speaking, it is a topological classification based on density distribution. Different from all previous method, a special resolution parameter is used in our model so that we can systematically organize the configuration states into various state with different level of scales. The multiscale property is achieved through our multiscale rigidity function.

\subsection{Multiscale rigidity function}\label{sec:Multiscale_rigidity_function}
The rigidity function is originally devised for the representation of biomolecular densities\cite{KLXia:2015c,KLXia:2015d}. It is a continuous version of rigidity index. In our previous works, we found that our flexibility and rigidity index can be very powerful in the study of biomolecular flexibility, particularly in the experimental b-factor prediction\cite{KLXia:2013d,Opron:2014,Opron:2015communication,Opron:2014b, Xia:2015multiscale,Nguyen:2016generalized}. Further, a rigidity functions with a special resolution parameter can deliver a multiscale representation of biomolecular structures\cite{KLXia:2015c,KLXia:2015d}. Through the variation of resolution parameter value, we can focus our geometric and topological ``lens" on the scales of interests. This special function is called multiscale rigidity function. A brief introduction is given below.

\begin{figure}
\begin{center}
\begin{tabular}{c}
\includegraphics[width=0.95\textwidth]{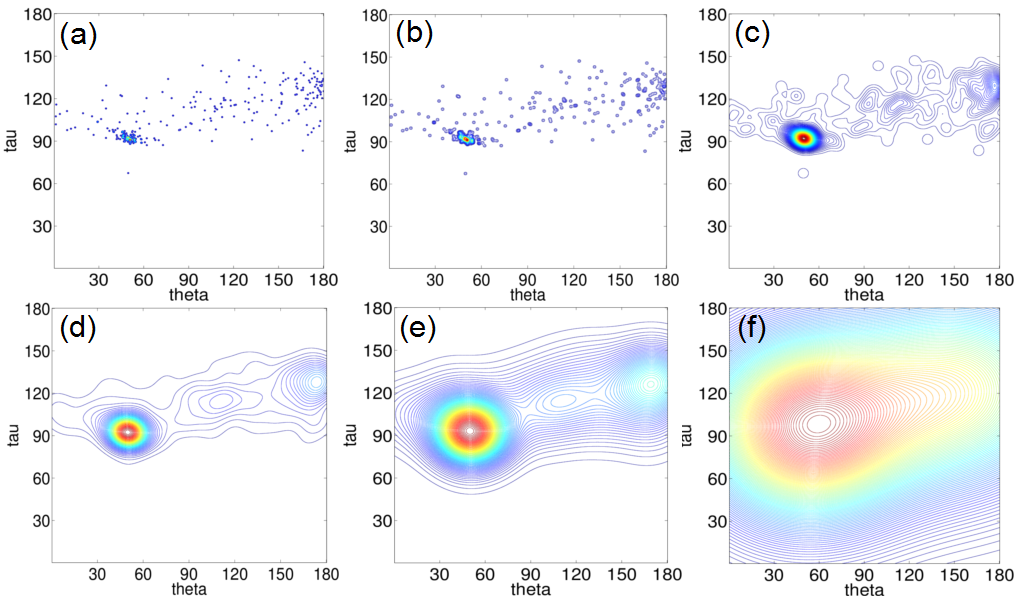}
\end{tabular}
\end{center}
\caption{The contour plots of the multiscale density functions for $(\theta, \tau)$ angles of protein 1VJU. The Gaussian kernel is used and values of scale parameter $\eta$ in (a)-(f) are $0.5^{\rm o}$, $1.0^{\rm o}$, $2.0^{\rm o}$, $10^{\rm o}$, $20.0^{\rm o}$ and $60.0^{\rm o}$, respectively.}
\label{fig:1vju_contour}
\end{figure}
For a data set with a total $N$ entries, which can be physical elements like atoms, residues and domains or data components like points, pixels and voxels, if we assume their generalized coordinates are ${\bf r}_1, {\bf r}_2,\cdots, {\bf r}_N$, a rigidity index of the $i$th entry can be expressed as
\begin{eqnarray}\label{eq:rigidity}
\mu_i=\sum_{j}^N w_j\Phi(r_{ij};\eta),
\end{eqnarray}
where $r_{ij}=\|{\bf r}_i-{\bf r}_j \|$ is the generalized distance between the $i$th and $j$th entries, $w_j$ is a weight and  $\Phi(r_{ij};\eta)$ is a kernel function satisfying the following admissibility conditions
\begin{eqnarray}\label{eq:couple_matrix1-1}
\Phi( r_{ij};\eta)&=&1 \quad {\rm as }\quad  r_{ij}   \rightarrow 0\\ \label{eq:couple_matrix1-2}
\Phi( r_{ij};\eta)&=&0 \quad {\rm as }\quad  r_{ij}   \rightarrow\infty.
\end{eqnarray}
Here $\eta>0$ is a resolution parameter that can be adjusted to achieve the desirable resolution for a given scale. Commonly used  correlation functions are  generalized exponential  functions
\begin{eqnarray}\label{eq:couple_matrix1}
\Phi( r_{ij};\eta, \kappa) =    e^{-\left( r_{ij} /\eta \right)^\kappa},    \quad \kappa >0
\end{eqnarray}
or  generalized Lorentz functions
\begin{eqnarray}\label{eq:couple_matrix2}
 \Phi( r_{ij};\eta, \upsilon) =  \frac{1}{1+ \left( r_{ij} /\eta \right)^{\upsilon}},  \quad  \upsilon >0.
 \end{eqnarray}
It can be noticed that the larger the $\eta$ value, the lower the resolution is.

The multiscale rigidity function of the data can be expressed as,
\begin{eqnarray}\label{eq:rigidity_function}
\mu({\bf r})=\sum_{j}^N w_j\Phi(\parallel {\bf r}- {\bf r}_{j}\parallel;\eta)
\end{eqnarray}

It can be seen that, with our rigidity function, a discrete data set is transformed into a continuous function. More importantly, the resolution  parameter in the multiscale rigidity function enables us to represent the data at the scale of interest. Figure \ref{fig:1vju_contour} demonstrates a series of contour maps for rigidity functions in various resolutions. The multiscale rigidity function is generated from $(\theta, \tau)$ angles of protein 1VJU. With a small resolution value, more local pattern can be observed in our density function. These local features gradually disappears with the increasing of $\eta$ value and more global type of features emerges. More interesting, the topology of density distribution reveals the clustering information within the data. Simply speaking, if more points are concentrated in a special region, it will form a higher ``peak" in the density function. If there is less or no points in a region, a lower ``valley" is formed in the density function. A topological representation of all these ``peaks" and ``valleys" will naturally capture the distribution features and provides a new way of clustering. Actually, the persistent barcodes generated from persistent homology analysis can quantitatively ``measure" these features in density distribution. And barcode based persistence functions, particularly the persistence entropy, provides an entropy estimation from a totally different perspective. To have a through understanding of our multiscale persistent homology analysis, a brief introduction of persistent homology is given in the next section.

\subsection{Persistent homology analysis}\label{sec:PHA}
Algebraic topology is very important mathematical tool for the study of global connectivity or invariants in the structure\cite{Hatcher:2001,Munkres:1984,ComputationalTopologyBook}. It explores topological properties with algebraic tools, like group, homology, etc. Persistent homology is a newly-invented algebraic topology method for structure characterization. Unlike the previous topological tools, it is able to define a geometric measurement for the generator of homology, thus it demonstrates great power for not only qualitative but also quantitative characterization of the structure. It plays a very important role in topological data analysis (TDA) and has been used in the analysis of biomolecular structure, flexibility and dynamics. A brief introduction of the essential element in persistent homology is given in this section.

\subsubsection{Simplicial homology}\label{sec:SimplicialHomology}
Simplicial complex is a finite set of simplices, which can be simply understood as vertices, edges, triangles, and their high dimensional counterparts. Simplicial complex, including geometric simplicial complex and abstract simplicial complex, is not a topological space, but it can be topologized as a subspace of $R^n$ called polyhedron. Groups and group operations can be defined on simplices. In this way, algebraic tools, particularly homology analysis, can be used to analyze the topological properties.

\paragraph{Simplicial complex}
Simplices are the build block for simplicial complex. A $k$-simplex is the convex hall of $k+1$ affine independent points in $\mathbb{R}^N$ ($N>k$). For a set of $k+1$ affine independent points $v_0,v_1,v_2,\cdots,v_k$, a correlated $k$-simplex $\sigma^k=\{v_0,v_1,v_2,\cdots,v_k\}$ can be expressed as
\begin{eqnarray}\label{eq:couple_matrix1}
\sigma^k=\left\{\lambda_0 v_0+\lambda_1 v_1+ \cdots +\lambda_k v_k \mid \sum^{k}_{i=0}\lambda_i=1;0\leq \lambda_i \leq 1,i=0,1, \cdots,k \right\}.
\end{eqnarray}
A geometric $k$-simplex $\sigma^k$ is a a closed convex compact connected subspace of $R^n$. Its $i$-dimensional face is the convex hall formed by $i+1$ vertices from $\sigma^k$ ($k>i$). Geometrically, a 0-simplex is a vertex, a 1-simplex is an edge, a 2-simplex is a triangle, and a 3-simplex represents a tetrahedron. We can also define the empty set as a (-1)-simplex.

A simplicial complex $K$ is a finite set of simplices that satisfy two essential conditions, i.e., 1) any face of a simplex from  $K$  is also in  $K$; 2) the intersection of any two simplices in  $K$ is either empty or  shared faces. The dimension of a simplicial complex is the maximal dimension of its simplicies. The polygon $|K|$ is a topological space formed by the union of all the simplices of $K$, i.e.,  $|K|=\cup_{\sigma^k\in K} \sigma^k$. In order to study the topological space with algebraic tools, we need to introduce the concept of chain.

\paragraph{Homology}
A linear combination $\sum^{k}_{i}\alpha_i\sigma^k_i$ of $k$-simplex $\sigma^k_i$ is known as $k$-chain $[\sigma^k]$. Coefficients $\alpha_i$ can be selected from various fields, including rational field $\mathbb{Q}$, integer field $\mathbb{Z}$, and prime integer field  $\mathbb{Z}_p$ with prime number $p$. In computational topology, coefficients $\alpha_i$ is chosen in the field of $\mathbb{Z}_2$ for simplicity. An Abelian group $C_k(K, \mathbb{Z}_2)$ is formed by the set of all $k$-chains from the simplicial complex $K$ together with addition operation. 

The boundary operation is essential to the definition of homology. A boundary operator $\partial_k$ is defined as $\partial_k: C_k \rightarrow C_{k-1}$. The boundary of a $k$-chain $[\sigma^k]=[v_0,v_1,v_2,\cdots,v_k]$ can be denoted as,
\begin{eqnarray}
\partial_k [\sigma^k] = \sum^{k}_{i=0} (-1)^i [ v_0, v_1, v_2, \cdots, \hat{v_i}, \cdots, v_k ].
\end{eqnarray}
Here $[v_0, v_1, v_2, \cdots ,\hat{v_i}, \cdots, v_k ]$ means a $(k-1)$ chain generated by the elimination of vertex $v_i$. Also we have $\partial_0= \emptyset$. From its definition, it can be found that if applying the boundary operation twice, any $k$-chain will be mapped to a zero element as $\partial_{k-1}\partial_k= \emptyset$. Further, the $k$-th cycle group $Z_k$ and the $k$-th boundary group $B_k$ are the subgroups of $C_k$ and can be defined as,
\begin{eqnarray}
&& Z_k={\rm Ker}~ \partial_k=\{c\in C_k \mid \partial_k c=\emptyset\}, \\
&&  B_k={\rm Im} ~\partial_{k+1}= \{ c\in C_k \mid \exists d \in C_{k+1}: c=\partial_{k+1} d\}.
\end{eqnarray}
Their elements are called the $k$-th cycle and the $k$-th boundary, respectively. It can be noticed that $B_k\subseteq Z_k \subseteq C_k$ as the boundary of a boundary is always empty $\partial_{k-1}\partial_k= \emptyset$.
The $k$-th homology group $H_k$ is the quotient group generated by the $k$-th cycle group $Z_k$ and $k$-th boundary group $B_k$: $H_k=Z_k/B_k$. The rank of $k$-th homology group is called $k$-th Betti number and it can be calculated by
\begin{eqnarray}
\beta_k = {\rm rank} ~H_k= {\rm rank }~ Z_k - {\rm rank}~ B_k.
\end{eqnarray}
Based on the fundamental theorem of finitely generated abelian group, homology group $H_k$ can be further expressed as a direct sum,
\begin{eqnarray}
H_k= {Z}\oplus \cdots \oplus {Z} \oplus {Z}_{p_1}\oplus \cdots \oplus {Z}_{p_n}= {Z}^{\beta_k} \oplus {Z}_{p_1}\oplus \cdots \oplus {Z}_{p_n},
\end{eqnarray}
where the rank of the free subgroup is the $k$-th Betti number $\beta_k$. Here $ {Z}_{p_i}$ is torsion group with torsion coefficients $\{p_i| i=1,2,...,n\}$.

Simply speaking, the geometric meanings of Betti numbers in $\mathbb{R}^3$ are the follows: $\beta_0$ represents the number of isolated components, $\beta_1$ is the number of one-dimensional loops, circles, or tunnels and $\beta_2$ describes the number of two-dimensional voids or holes. 

\paragraph{$\check{\rm C}$ech complex, Rips complex and alpha complex}
There are various ways of defining simplicial complices for a given topological space including $\check{\rm C}$ech complex, Rips complex and alpha complex. All these definitions are developed from a simple idea called Nerve. In topology, the nerve of an an open covering delivers an abstract simplicial complex from the open covering of a topological space. More specifically, given an index set $I$ and open set ${\bf U}=\{U_i\}_{i\in I}$ which is a open covering  of  a topological space $X \in \mathbb{R}^N$, i.e., $X \subseteq \{U_i\}_{i\in I}$, the nerve {\bf N} of {\bf U} satisfies two basic conditions, i.e., 1) $\emptyset \in {\bf N}$; 2) if $\cap_{j \in J} U_j \neq \emptyset $ for $J \subseteq I $, one has $J \in {\bf N}$.

$\check{\rm C}$ech complex is the nerve of a cover constructed from the union of balls. For a point set $X \in \mathbb{ R}^N$, one defines a cover of closed balls ${\bf B}=\{B (x,\epsilon)\mid x \in X \}$ centered at $x$ with radius $\epsilon$. The $\check{\rm C}$ech complex of $X$ with parameter $\epsilon$ is denoted as $\mathcal{C}(X,\epsilon)$, which is the  nerve of the closed ball set {\bf B},
\begin{eqnarray}
\mathcal{C}(X,\epsilon) = \left\{ \sigma \mid \cap_{x \in \sigma} B (x,\epsilon) \neq \emptyset \right\}.
\end{eqnarray}
The condition for a $\check{\rm C}$ech complex can be relaxed to generate a Vietoris-Rips complex (or Rips complex) denoted as $\mathcal{R}(X,\epsilon)$ \cite{Edelsbrunner:1994}. In Rips complex, a simplex $\sigma$ will be generated if the largest distance between any of its vertices reaches $2\epsilon$. These two abstract complexes satisfy the relation,
\begin{eqnarray}\label{eq:SandwichRelation}
\mathcal{C}(X,\epsilon)\subset \mathcal{R}(X,\epsilon) \subset \mathcal{C}(X,\sqrt{2}\epsilon).
\end{eqnarray}

Derived from computational geometry, alpha complex is also an important geometric concept. It is derived from Voronoi diagram and Delaunay complex. Let $X$ be a point set in Euclidean space $\mathbb{R}^d$. The Voronoi cell of a point $x \in X$ is defined as
\begin{eqnarray}
V_x = \{ u\in R^d \mid |u-x|\leq |u-x'|, \forall x'\in X \}.
\end{eqnarray}
The collection of all Voronoi cells forms a Voronoi diagram and the nerve of the Voronoi diagram generates a Delaunay complex. Further, if we define $R(x,\epsilon)$ as the intersection of Voronoi cell $V_x$ with ball $B(x,\epsilon)$, i.e., $R(x,\epsilon)= V_x \cap B(x,\epsilon)$, the alpha complex $\mathcal{A}(X,\epsilon)$ of point set $X$ is just the nerve of cover $\cup_{x\in X} R(x,\epsilon)$,
\begin{eqnarray}
\mathcal{A}(X,\epsilon) = \left\{ \sigma \mid \cap_{x \in \sigma} R (x,\epsilon) \neq \emptyset \right\}.
\end{eqnarray}
It can be seen that an alpha complex is a subset of a Delaunay complex.

\subsubsection{Persistent homology analysis}\label{sec:ConstructionHomology}
As stated above, a radius parameter $\epsilon$ is consistently used in the open covering to recover the original topological space. However, how to find the best suitable $\epsilon$ for the underling space has puzzled researchers for a long time. It is true that when $\epsilon$ is too small, originally connected regions may not be fully recovered. But when $\epsilon$ is too large, originally non-connected regions may be mistaken as connected. To solve this problem, a brilliant idea has been proposed and it is known as filtration. In the filtration process, through a systematical investigation of a wide range of $\epsilon$ values, a series of topological spaces from various scales have been generated. Their topological information can be evaluated. It is found that some topological invariants last for several scales or a large range of $\epsilon$ values, and some invariants disappear very quickly when scale changes or  $\epsilon$ value changes. State different, all these topological invariants can be granted with a "lifespan" or persistence. This means that a special geometric measurement (range of $\epsilon$) can be found for each topological invariant. This method is known as the persistent homology. It differs greatly from the traditional geometric and topological methods by the incorporation of geometric information into topological invariants. It can work as a bridge between geometry and topology \cite{Bubenik:2007, edelsbrunner:2010,Dey:2008,Dey:2013,Mischaikow:2013}.
\paragraph{General filtration processes}
Filtration is of essential importance to persistent homology. A suitable filtration is key to the persistent homology analysis. In practice, two filtration algorithms, Euclidean-distance based and density based ones, are commonly used.  These filtration processes can be modified in many different ways to address physical needs as shown in our previous papers 

The basic Euclidean-distance based filtration is straightforward. One associates each point with an ever-increasing  radius to form an ever-growing ball for each point. When these balls gradually overlap with each other, complexes can be identified from various complex construction algorithms described above. In this manner, the previously formed simplicial complex is an inclusion of latter ones and  naturally, a filtration process is created. One can  formalize this process by  the use of a distance matrix $\{d_{ij}\}$. Here the matrix element $d_{ij}$ represents the distance between atom $i$ and atom $j$. For diagonal terms, it is nature to assume $d_{ij}=0$.  Let us  denote the filtration threshold as a parameter $\varepsilon$. A 1-simplex is generated between vertices $i$ and $j$ if $d_{ij}\leq\varepsilon$. Similarly higher dimensional complexes can also be defined.


Another important filtration process is the density based filtration process. In this process, the filtration goes along the increase or decrease of the density value. In this way, a series of isosurfaces are generated. Morse complex is used for the characterization of their topological invariants. Persistence information can be derived from these complices. A more rigorous mathematical formulation is given in the following.

\paragraph{Persistent homology}
The filtration can be described as a nested sub-sequence of its subcomplexes,
\begin{eqnarray}
\varnothing = K^0 \subseteq K^1 \subseteq \cdots \subseteq K^m=K.
\end{eqnarray}
Generally speaking, abstract simplicial complexes generated from a filtration give a multiscale representation of the corresponding topological space, from which related homology groups can be evaluated to reveal  topological features. Furthermore, the $p$-persistent $k$-th homology group $K^i$ can be represented as
\begin{eqnarray}
H^{i,p}_k=Z^i_k/(B_k^{i+p}\bigcap Z^i_k).
\end{eqnarray}
Through the study of the persistent pattern of these topological features, the persistent homology is capable of capturing the intrinsic properties of the underlying space.

To visualize the persistent homology results, many elegant representation methods have been proposed, including persistent diagram\cite{Mischaikow:2013}, persistent barcode\cite{Ghrist:2008barcodes}, persistent landscape\cite{Bubenik:2015statistical}, etc. In this paper, we use a barcode representation. A simple example can be found in Figure \ref{fig:1vju_sig8} (c). Basically, barcodes are an assemble of bars. Each bar represents a homology generator with ``birth" and ``death" time as its starting and ending points. In this way, the length of bar tells how long the homology generator ``lives" or ``persists".

\subsection{Multiscale persistent function}\label{sec:MPH}
\begin{figure}
\begin{center}
\begin{tabular}{c}
\includegraphics[width=0.95\textwidth]{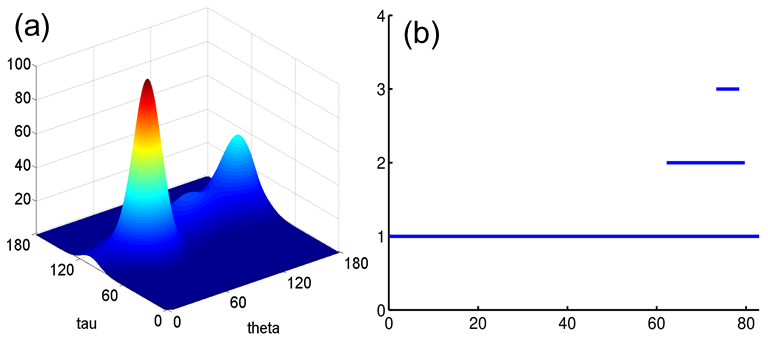}
\end{tabular}
\end{center}
\caption{The illustration of rigidity function and density filtration of $(\theta, \tau)$ angles of protein 1VJU (chain A). (a) Rigidity function constructed by using Gaussian kernel with a scale parameter $\eta=8$\AA. (b) The barcodes representation of density filtration for protein 1VJU (chain A). The x-axis is the rescaled density value. Three bars corresponded to the three peaks in the density map and contour plot can be observed. And in this way, the points in Figure \ref{fig:1vju_psi_phi} (c) is naturally subdivided into three regions represented by three individual bars in barcodes. And the probability for each subdomain equals to the ratio between its barlength and the total barlength.}
\label{fig:1vju_sig8}
\end{figure}
With the brief introduction of multiscale rigidity function and persistent homology, now we can introduce our multiscale persistent homology and multiscale persistent functions. Simply speaking, for each resolution, a density filtration over the rigidity function is performed. We let the filtration goes from the highest density value to the lowest one. A series of barcodes can be generated and represented as following,
\begin{eqnarray}
\{ B_{k,j}(\eta)=[a_{k,j}(\eta), b_{k,j}(\eta)] | k=0,1,2,...; j=1,2,3,....,N_k \}.
\end{eqnarray}
Here parameter $k$ is the dimension of the homology group $\beta_k$, parameter $j$ indicates the $j$-th barcode and $N_k$ is the number of $\beta_k$ barcodes. 
Barcodes can be employed to construct various persistent functions.  In our previous works\cite{KLXia:2014persistent,KLXia:2015a}, we have defined the accumulated barlength,
\begin{eqnarray}
A_b(B_{k,j})=\sum_j (b_{k,j}-a_{k,j}),
\end{eqnarray}
and use it to study the bond energy and folding energy in protein unfolding simulation. In the fullerene total curvature energy evaluation, the intrinsic barcode of $\beta_2$\cite{KLXia:2015a}
\begin{eqnarray}
I_b(B_{2,j})=max((b_{2,j}-a_{2,j})),
\end{eqnarray}
is also found to have important physical meaning.

In this section, based on the multiscale persistent homology results, we propose several types of multiscale persistent energy functions. The first type is

\begin{eqnarray}
f_1(x;B_{k,j}(\eta))= \sum_{j} w_{k,j} e^{-\left(\frac{x- (b_{k,j}(\eta)+a_{k,j}(\eta))/{2}}{\sigma(\eta) (b_{k,j}(\eta)-a_{k,j}(\eta))}\right)^\kappa},  \quad \kappa >0
\end{eqnarray}
here $w_{m,j}$ is the weight function for the $j$-th barcode of $\beta_m$ and the parameter $\sigma(\eta)$ is characterization of ``significance" of barcode in different scales.  The second important type of persistent energy functions is
\begin{eqnarray}
f_2(x;B_{k,j}(\eta))= \sum_{j} w_{k,j} \frac{1}{1+ \left(\frac{x-(b_{k,j}(\eta)+a_{k,j}(\eta))/{2}} {\sigma(\eta) (b_{k,j}(\eta)-a_{k,j}(\eta))}\right)^\upsilon}, \quad  \upsilon >0.
\end{eqnarray}
It can be noticed that this two multiscale function can be used to visualize the barcodes information. And more importantly, all these persistent energy functions can be used to analyze the molecular dynamic trajectory and study the multiscale information in the data.

\paragraph{Multiscale persistent entropy}

\begin{figure}
\begin{center}
\begin{tabular}{c}
\includegraphics[width=0.95\textwidth]{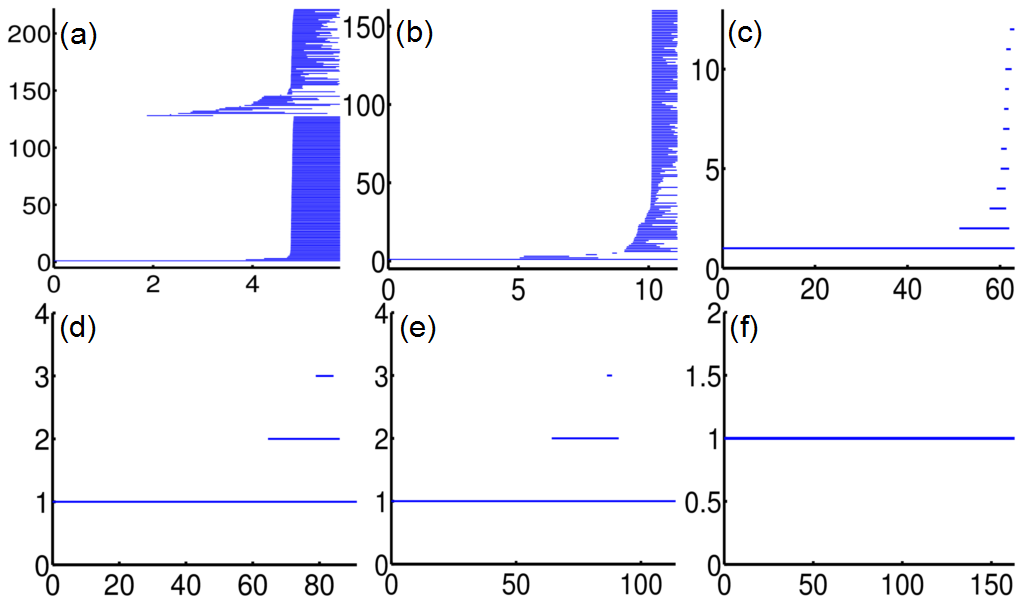}
\end{tabular}
\end{center}
\caption{The barcodes of the multiscale density function derived from protein 1VJU (chain A) $\theta-\tau$ data. The Gaussian kernel is used and values of scale parameter $\eta$ in (a)-(f) are $0.5^{\rm o}$, $1.0^{\rm o}$, $2.0^{\rm o}$, $10^{\rm o}$, $20.0^{\rm o}$ and $60.0^{\rm o}$, respectively.}
\label{fig:1vju_barcodes}
\end{figure}

More interesting, we can define a multiscale persistent entropy as following:
\begin{eqnarray}\label{eq:filtrationM}
S_k(\eta)=\sum_j^{N_k(\eta)} - p_{j,k}(\eta) log(p_{j,k}(\eta)),
\end{eqnarray}
with the probability function
\begin{eqnarray}\label{eq:pi}
p_{k,j}(\eta)=\frac{b_{k,j}(\eta)-a_{k,j}(\eta)}{\sum_j (b_{k,j}(\eta)-a_{k,j}(\eta))}.
\end{eqnarray}
The multiscale persistent entropy can also be simplified as following:
\begin{eqnarray}\label{eq:MPE}
S_k(\eta)=log\left(\sum_j^{N_k(\eta)} (b_{k,j}(\eta)-a_{k,j}(\eta))\right)- \frac{\sum_j^{N_k(\eta)}\left( (b_{k,j}(\eta)-a_{k,j}(\eta)) log(b_{k,j}(\eta)-a_{k,j}(\eta)) \right)}{\sum_j^{N_k(\eta)} (b_{k,j}(\eta)-a_{k,j}(\eta))},
\end{eqnarray}

It can be seen that, based on the density filtration, $\beta_0$ bars are able to capture the topological connectivity of the density map and naturally incorporate in them the information of clustering. And persistent entropy $S_0(\eta)$ provides a measurement of intrinsic disorderliness in the data. To illustrate this idea more clearly, we analyze the $(\theta, \tau)$ angles of protein 1VJU (chain A) in Figure \ref{fig:1vju_psi_phi} (c). Through a Gaussian kernel with resolution parameter $\eta=8$\AA~, a density distribution is generated in Figure \ref{fig:1vju_sig8}(a). We can observe three ``peaks" in the density map, meaning the original data points are concentrated in three clusters in this resolution. Further, larger clusters result in higher ``peaks", that is to say the size of clusters can be measured by the relative heights of the ``peaks". All these information is naturally incorporated into our persistent barcods illustrated in Figure \ref{fig:1vju_sig8}(b). Three bars representing three clusters have different lengths and length measures the density or point numbers in the cluster. Essentially, each $\beta_0$ is a cluster in the data and the barlength represents the size of the cluster. In this way, multiscale persistent entropy gives a natural measurement of the disorderliness of the data.

More importantly, with various resolution values, the barcodes results are able to give a full ``spectrum" information of the data. Figure \ref{fig:1vju_barcodes} illustrates multiscale barcodes for $(\theta, \tau)$ angles of protein 1VJU. The $(\theta, \tau)$ angle based multiscale rigidity functions are generated with scale parameter $\eta=0.5^{\rm o}$, $1.0^{\rm o}$, $2.0^{\rm o}$, $10^{\rm o}$, $20.0^{\rm o}$ and $60.0^{\rm o}$ as indicated in Figure \ref{fig:1vju_contour}. From the comparison of the density function contours and barcode results, it can be seen that our $\beta_0$ bars capture the topological features of the density map very well. When resolution value is small, local details of the density maps are revealed. So more short barcodes emerge. And correspondingly, we classify data into more clusters. When resolution value is large, local details are smoothed with only a few long persisting bars in the barcodes. In this situation, data is classified into only a few clusters. Depending on the scale we interested, the resolution values can be systematically adjusted. So are the classification clusters.

Unlike the traditional regular mesh discretization, which employs an uniformed regular mesh, our discretization and classification is highly dependent on the data structure, thus preserving more topological features in the original data. To have a clear picture of the difference between the two approaches, we compare the our multiscale persistent entropy with the general conformational entropy evaluation. Again we use the same data set with 110 proteins as in Section \ref{sec:conformation_entropy} and the protein IDs are listed in Table \ref{tab:protein_index}.

\begin{figure}
\begin{center}
\begin{tabular}{c}
\includegraphics[width=0.9\textwidth]{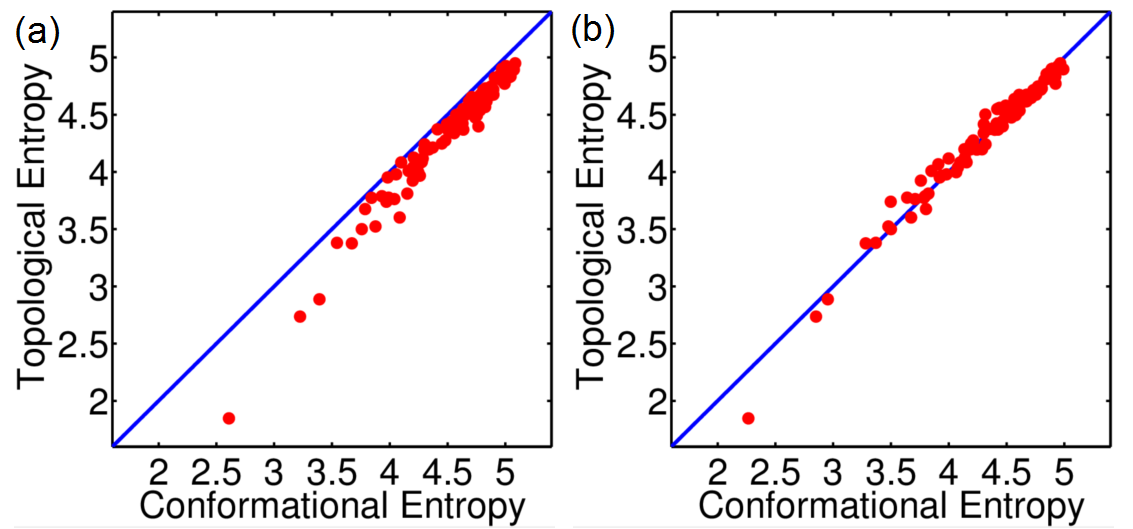}
\end{tabular}
\end{center}
\caption{The comparison of persistent entropy at high resolution with traditional entropy at refined mesh. The scale parameter $\eta$ is $1.0^{\rm o}$. The grid spacings are $2.0^{\rm o}$ and $3.0^{\rm o}$ in (a) and (b). The PCC is 0.984 and 0.988 in (a) and (b).
}
\label{fig:comparison_sig1}
\end{figure}

Generally speaking, our scale parameter $\eta$ is a counterpart of grid spacing size in the traditional conformation entropy method. Since the Gaussian kernel has a relative dominant range about $3\eta$, we choose the size of the grad spacing to be about 2 to 3 times of $\eta$ and make a comparison. It is seen that when grid spacing is very small, all the evaluated entropy values approach to $ln(N_{res}-3)$ as indicated in Figure \ref{fig:entropy_grid}. This property is well-preserved in our scheme. As it can be seen that there is a very large correlation for the small $\eta$ value as demonstrated in Figure \ref{fig:comparison_sig1} ({\bf a}). Further, when the grid spacing values are extremely large, all entropy values approach to zero in the traditional methods as demonstrated in Figure \ref{fig:entropy_grid}({\bf h}). This is also true in our scheme. When $\eta$ value is very large, density map can have only one peak as indicated in Figure \ref{fig:1vju_contour}({\bf f}) and the corresponding barcodes have only one long persisting bar as illustrated in Figure \ref{fig:1vju_contour}({\bf f}). In this way, our persistent entropy equal exactly to zero just as the same as the traditional methods. And a high correlation can also been achieved.

\begin{figure}
\begin{center}
\begin{tabular}{c}
\includegraphics[width=0.9\textwidth]{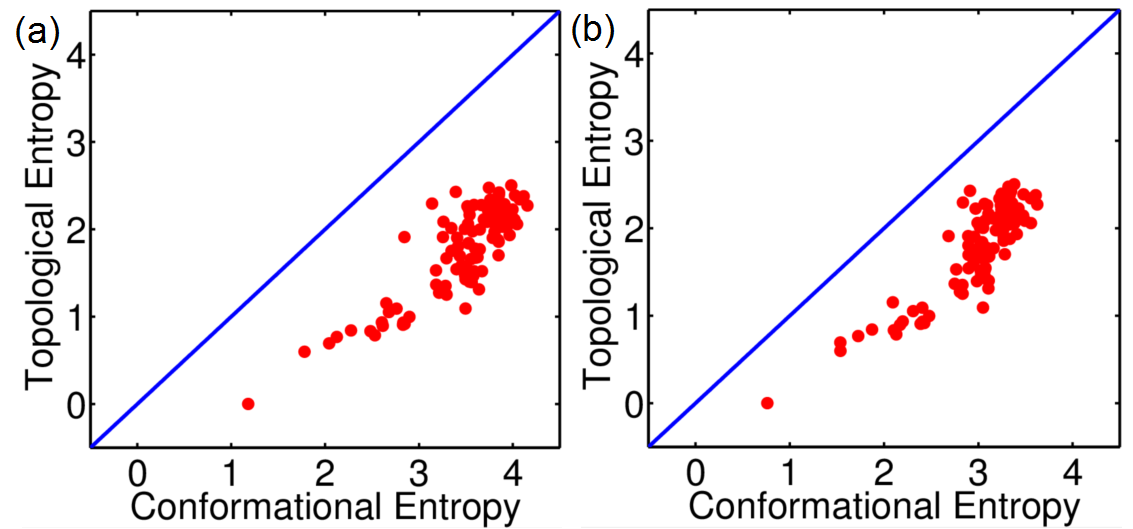}
\end{tabular}
\end{center}
\caption{The comparison of persistent entropy at medial resolution with traditional entropy at medially refined mesh.
The scale parameter $\eta$ is $5.0^{\rm o}$. The grid spacings are $10.0^{\rm o}$ and $15.0^{\rm o}$ in (a) and (b). The PCC is 0.848 and 0.863 in (a) and (b).
}
\label{fig:comparison_sig5}
\end{figure}

More interesting is the situation when the value of scale parameter and grid spacing values are in the middle range. In this situation, two methods do deliver different results. This is due to the reason that different ways are employed for data classification. Figure \ref{fig:comparison_sig5} demonstrates this differences. The Pearson correlation coefficients decrease greatly compared with the small scale situation. In the traditional conformation entropy method, it has already been suggested that the grid spacing is chosen in the range around $10^{\rm o}$ to $60^{\rm o}$. And this is exactly the range in which our classification is dramatically different from the traditional ones.

\section{Application} \label{sec:application}
In this section, two examples are used to showcase our multiscale persistent homology analysis, particularly the multiscale persistent entropy. The first case studied is protein classification. Based on our entropy values, we can have a nice classification of all-alpha, all-beta and mixed alpha and beta proteins. Especially for all-alpha and all-beta proteins, we can have a clearly separation in their entropy values. Mixed-alpha-and-beta proteins have entropy values largely in the gap between all-alpha and all-beta ones. And based on this results, we propose a protein structure index (PSI). The PSI can be used to measure the regularity of protein structures. Simply speaking, it provides a way to characterize the disorderliness in the system. This index can be applied to any other data types.

\subsection{The protein classification}\label{sec:classification}
\begin{figure}
\begin{center}
\begin{tabular}{c}
\includegraphics[width=0.6\textwidth]{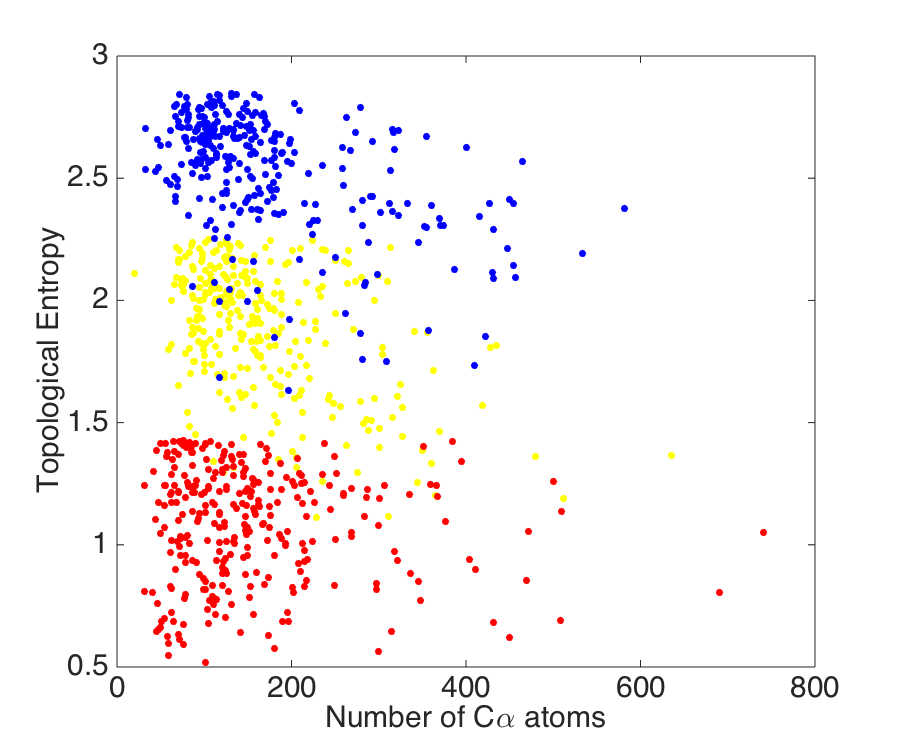}
\end{tabular}
\end{center}
\caption{The persistent entropy based protein classification. The red, blue, and yellow points are from all-alpha, all-beta and mixed-alpha-and-beta proteins.}
\label{fig:entropyVSAtom_scatter}
\end{figure}

Based on secondary structure properties, proteins can be classified into three categories, including all-alpha, all-beta, or mixed-alpha-and-beta proteins. In general, two common motifs of secondary structure, i.e.,the $\alpha$-helix and the $\beta$-sheets, are defined by hydrogen bond patterns of the main-chain peptide. Roughly speaking, if protein backbone form a helix structure, i.e., a spiral conformation, this special region is called $\alpha$-helix; if the backbone form a twisted or pleated sheet, the backbone part is known as $\beta$-sheets. Both the $\alpha$-helix and the $\beta$-sheet represent a way of saturating all the hydrogen bond donors and acceptors in the peptide backbone.

In this section, persistent entropy is employed for the classification of protein structures. We employ the test dataset used in \cite{ZXCang:2015} and originally downloaded from SCOPe (Structural Classification of Proteins-extended) database. In this test case, three types of proteins are selected and for each class, 300 structures from different superfamilies are chosen. In our approach, for each protein, we calculate ($\theta$, $\tau$) angles first and then evaluate its persistent entropy using a Gaussian kernel with scale parameter $\eta=5^{\rm o}$.  Figure \ref{fig:entropyVSAtom_scatter} shows our persistent entropy results.  The red, blue, and yellow points are from all-alpha, all-beta and mixed-alpha-and-beta samples, respectively. The values of persistent entropy from all-alpha are systematically lower than those from all-beta sample. And there is clear interface between these two samples. It infers that persistent entropy can be used as a good candidate to classify all-alpha and all-beta proteins. In the mean time, the persistent entropies from the mixed-alpha-and-beta sample are in the middle between all-alpha and all-beta as expected.

It can be seen that ($\theta$, $\tau$) angle based persistent entropy provides an efficient approach to qualitatively classify the protein structures. More interesting, this unique persistent entropy provides a quantitative model for the characterization of protein structure ``regularity". We call ($\theta$, $\tau$) angle based persistent entropy as protein structure index (PSI).

\subsection{Protein structure index}\label{sec:protein_index}
\begin{figure}
\begin{center}
\begin{tabular}{c}
\includegraphics[width=0.9\textwidth]{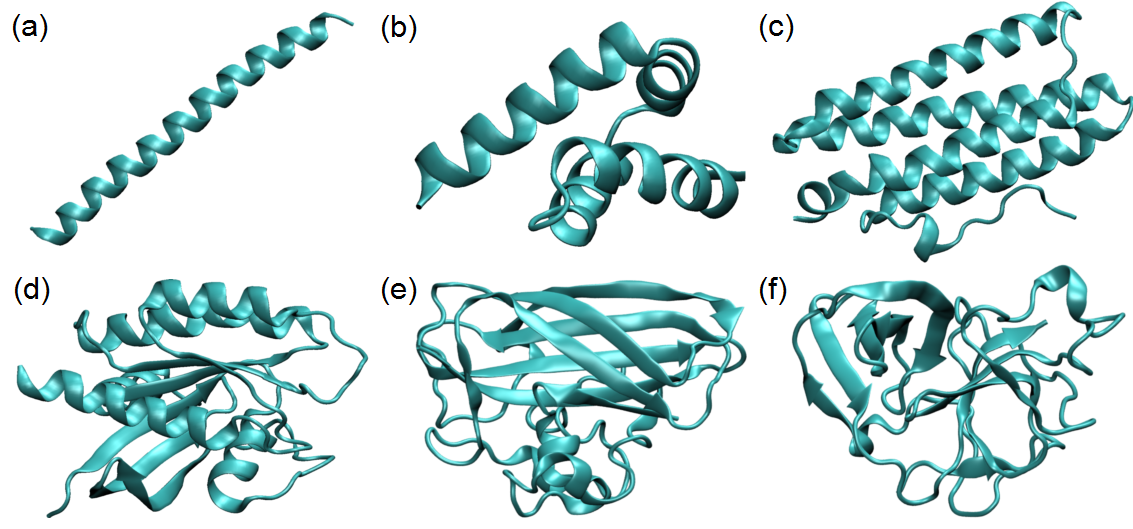}
\end{tabular}
\end{center}
\caption{The protein structure index derived from persistent entropy. The proteins are 1GK7(0.00), 1I2T(0.60), 4XQI(0.84), 4EPV(1.52), 4ALT (2.11), and 3ZFP(2.50)}
\label{fig:protein_index}
\end{figure}

The essential idea of protein structure index is to evaluate the degree of disorderliness in protein ($\theta$, $\tau$) angle data. To be more specific, we find that when the protein structure contains only $\alpha$-helix, its ($\theta$, $\tau$) plot is very regular with data points highly concentrated around $(50^{\rm o}, 90^{\rm o})$ and the evaluated persistent entropy is very low. The reason is that $\alpha$-helix is a highly ``regular" spiral conformation. Similarly, $\beta$-sheet is also very regular structure with ($\theta$, $\tau$) angles concentrated around $(195^{\rm o}, 117^{\rm o})$. However, $\beta$-sheet has various twist configurations so that the ($\theta$, $\tau$) angles tend to more scattering than the ones in $\alpha$-helix. Persistent entropy for $\beta$-sheet is generally larger than that of $\beta$-sheet. Further, if protein structures have a large portion of loops or intrinsic disordered regions, ($\theta$, $\tau$) angles will be more diverse and the correspondingly persistent entropy will be even higher. Physically, loops and intrinsic disordered regions are usually very unstable compared with $\alpha$-helix and $\beta$-sheet. Generally, loop regions have large B-factors and can be very flexible. Intrinsic disorder regions even lack a fixed or ordered three-dimensional structure. In this way, ($\theta$, $\tau$) angle based persistent entropy provides a quantitative characterization of protein structure regularity.

To evaluate our PSI model, we choose 110 proteins with resolution higher or equal to 1.5\AA. The protein ID are listed in  Table \ref{tab:protein_index}. Most the structures have only one chain in it except eight structures marked by star. For these data, we remove all the other chains from it leaving only the first chain (chain A) in the structure. In this way, there is no unphysical dihedral and bond angles in our ($\theta$, $\tau$) plot due to the dislocation between the ends of two chains. We then systematically calculate the two types of angles ($\theta$, $\tau$), transform the point cloud data into density representation and employ the persistent homology analysis. We use Gaussian kernel with scale parameter $\eta=5^{\rm o}$. We have illustrated the results in Figure \ref{fig:protein_index}. It can be seen that the protein structure index provides a comparably nice description of protein regularity. The smallest index indicates the most regular structure, i.e., a regular $\beta$-sheet. As the increase of the index, the protein structure gets more and more ``irregular". Again, by ``irregular", we mean large portion of loops and intrinsic disorder regions.

Our structure index can also be used to describe the regularity in any systems. Essentially, it provides a unique way of characterization the disorderliness in any structure or data. It should noticed that, our protein structure index is only based on the $\beta_0$ topological entropy. More interesting information can be derived from $\beta_1$ and $\beta_2$ topological entropies. This however will not be discussed in the current paper.

\begin{table}[htbp]
  \centering
	\caption{Protein structure index (PSI) for 110 proteins, the topological entropy is evaluated by using Gaussian kernel with scale parameter $\eta=5^{\rm o}$. Protein with a smaller index value is more regular. Protein with larger index usually has more loops or disorder regions.}
 \begin{tabular}{||cc||cc||cc||cc||cc||} \hline \hline
 Protein ID  &PSI  &Protein ID  &PSI &Protein ID  &PSI &Protein ID  &PSI &Protein ID  &PSI \\ \hline
 1GK7 &0.000  &1I2T  &0.599  &1TQG  &0.694  &3BT5  &0.768  &1JL6  &0.786 \\
 $2CB8^*$ &0.833  &4XQ1  &0.842  &2IMT  &0.895  &2F1S  &0.907  &2BBR  &0.917\\
 4W59 &0.932  &3SP7  &0.934  &2F2Q  &0.997  &3S0D  &1.053  &4DUI  &1.092 \\
 4WOH &1.093  &1LWB  &1.154  &4ETN  &1.252  &1IO0  &1.275  &2CLC  &1.312 \\
 $4J20^*$ &1.353  &4LD1  &1.365  &3DZE  &1.395  &2GKP  &1.401  &4NGJ  &1.429 \\
 4PZZ &1.451  &4DLR  &1.466  &1XEO  &1.473  &4EPV  &1.518  &3EYE  &1.520 \\
 3X1X &1.520  &2VIM  &1.530  &3L42  &1.543  &4LDJ  &1.545  &4B0D  &1.587 \\
 1R2Q &1.660  &3LF5  &1.669  &4J4Z  &1.675  &1RCF  &1.681  &4TKJ  &1.696 \\
 1KMV &1.703  &1SMU  &1.754  &3T3L  &1.756  &2NN5  &1.772  &4ICI  &1.778 \\
 3CCD &1.804  &4FXL  &1.840  &4BPY  &1.841  &4D0Q  &1.858  &3NXO  &1.880 \\
 2I24 &1.902  &$4HIL^*$  &1.903  &$4J46^*$  &1.911  &3U7T  &1.911  &4BJ0  &1.934 \\
 1TT8 &1.971  &2V1M  &1.978  &1TG0  &1.997  &2D4J  &2.005  &$2GBJ^*$  &2.015 \\
 4QT2 &2.029  &2C01  &2.043  &1IKJ  &2.050  &1ZK5  &2.053  &4DT4  &2.057 \\
 3R54 &2.060  &1OD3  &2.061  &4WDC  &2.062  &4IL7  &2.062  &4WEE  &2.066 \\
 2QDW &2.085  &4ACJ  &2.085  &2GKT  &2.087  &3EY6  &2.102  &4ALT  &2.110 \\
 2PND &2.114  &4NI6  &2.120  &2BSC  &2.132  &4CST  &2.133  &4JGL  &2.139 \\
 2ALL &2.142  &4ALR  &2.145  &3KZD  &2.166  &2XOM  &2.186  &4P5R  &2.213 \\
 $4TN9^*$ &2.224  &2Y6H  &2.228  &1KT7  &2.230  &1T2I  &2.263  &3M0U  &2.264 \\
 1XT5 &2.267  &4N1S  &2.272  &1HK0  &2.274  &4B9P  &2.275  &1ZIR  &2.275 \\
 2FT7 &2.278  &$2Y72^*$  &2.280  &2EU7  &2.287  &4JFM  &2.292  &2ON8  &2.295 \\
 4R4T &2.340  &4ZSD  &2.343  &4N1M  &2.379  &4DPC  &2.383  &1MFM  &2.388 \\
 1OH4 &2.395  &2J1A  &2.421  &$3NEQ^*$  &2.428  &3LL1  &2.476  &3ZFP  &2.504 \\ \hline \hline
    \end{tabular}%
  \label{tab:protein_index}
\end{table}

\section{Conclusion remarks}
In this paper, we discuss our multiscale persistent homology analysis, particulary multiscale persistent functions. The multiscale persistent homology analysis is based on two methods, i.e., multiscale rigidity function and persistent homology. The rigidity function is essential to a continuous version of rigidity index. It incorporate in it the multiscale information by turning a resolution parameter in kernel functions. Further, multiscale barcode representation of the data can be achieved by a density filtration over these multiscale rigidity functions. Multiscale persistent functions can be defined on these barcode spaces. We discuss a particulary function, i.e., persistent entropy, and illustrate its applications in protein structure classification and protein structure index representation.

There are several distinguished characteristics of our multiscale persistent entropy. Firstly, we naturally divide the data into several groups. This subdivision is based on the general topological features of the elements,i.e., ($\theta$, $\tau$) angle points. Secondly, the classification information is embedded into our $\beta_0$ barcodes. Essentially, each $\beta_0$ is a cluster in the data and the barlength represents the size of the cluster. It needs to be pointed out that our barcode based clustering is comparable with other methods, such as, K-mean methods\cite{Zhang:2008RNA}, spectral graph theory\cite{ChungOverview}, etc. In this sense, the topological entropy obtained in some special kernels and scales, can be similar to the results from the other clustering methods. However, topological entropies defined from high dimensional $\beta$ barcodes are dramatically different from all the previous methods. It should be noticed that topological invariants are able to describe the global structure information, and only persistent homology is able to characterize topological invariant with a geometric measurement, i.e., persistence. In this way, high dimensional barcode based topological entropies will be able to provide more interesting intrinsic topological information of the structure. This will be our future research topics.


\section*{Acknowledgments}

This work was supported in part by NTU SUG-M4081842.110. The author Zhiming Li thanks the Chinese Scholarship Council for the financial support No. 201506775038.

\vspace{0.6cm}


\end{document}